\documentclass[sigconf]{acmart}

\usepackage{balance}

\AtBeginDocument{%
  }

\setcopyright{acmlicensed}
\copyrightyear{2026}
\acmYear{2026}
\acmDOI{XXXXXXX.XXXXXXX}

\acmConference[CHIIR '26]{ACM SIGIR Conference on
Human Information Interaction and Retrieval}{March 22--26,
  2026}{Seattle, WA, USA}

\acmISBN{978-1-4503-XXXX-X/2018/06}

\begin{document}

\title{Privacy Discourse and Emotional Dynamics in Mental Health Information Interaction on Reddit}

\author{Jai Kruthunz Naveen Kumar}
\affiliation{
  \institution{George Mason University}
  \city{Fairfax, Virginia}
  \country{USA}
}
\email{jnaveenk@gmu.edu}

\author{Aishwarya Umeshkumar Surani}
\affiliation{
  \institution{Medical group management association (MGMA)}
  \city{Denver, Colorado}
  \country{USA}
}
\email{asurani@mgma.org}

\author{Harkirat Singh}
\affiliation{
\institution{George Mason University}
  \city{Fairfax, Virginia}
  \country{USA}
}
\email{hsingh45@gmu.edu}

\author{Sanchari Das}
\affiliation{
  \institution{George Mason University}
  \city{Fairfax, Virginia}
  \country{USA}
}
\email{sdas35@gmu.edu}

\begin{abstract}
Reddit is a major venue for mental-health information interaction and peer support, where privacy concerns increasingly surface in user discourse. Thus, we analyze privacy-related discussions across $14$ mental-health and regulatory subreddits, comprising $10{,}119$ posts and $65{,}385$ comments collected with a custom web scraper. Using lexicon-based sentiment analysis, we quantify emotional alignment between communities via cosine similarity of sentiment distributions, observing high similarity for Bipolar and ADHD ($0.877$), Anxiety and Depression ($0.849$), and MentalHealthSupport and MentalIllness ($0.989$) subreddits. We also construct keyword dictionaries to tag privacy-related themes (e.g., HIPAA, GDPR) and perform temporal analysis from $2020$ to $2025$, finding a $50\%$ increase in privacy discourse with intermittent regulatory spikes. A chi-square test of independence across subreddit domains indicates significant distributional differences ($\chi^2=5596.67$, $p=0.03$, $df=50$). The results characterize how privacy-oriented discussion co-varies with user sentiment in online mental-health communities.
\end{abstract}



\maketitle

\section{Introduction}
\label{sec:intro}

Social media platforms have redefined digital communication, connecting over $4.9$ billion users worldwide and projected to reach $5.8$ billion by $2027$~\cite{ForbesSocialMediaStats}. Reddit is a high-engagement, open-access platform with more than $100{,}000$ topic-specific subreddits and approximately $50$ million daily active users~\cite{kou2018understanding,cho2021potential,gruzd2020coding,kyto2017augmenting,das2020risk,sharevski2025social}, enabling peer-driven, real-time discussions that reflect collective sentiment and information exchange~\cite{li2023sentiment,proferes2021studying,oak2025victims,tally2023tips,gallagher2019reclaiming}. It also serves as a digital support ecosystem for mental health~\cite{kim2023understanding}, where individuals share lived experiences with depression, anxiety, and therapy~\cite{de2014mental,gupta2024critical,saha2020understanding,gupta2024really}. However, this openness raises concerns about privacy, anonymity, and the ethical handling of sensitive disclosures~\cite{boettcher2024case,yin2024navigating,chancellor2019taxonomy}, and prior work has documented risks from scraping and re-identification in online mental health spaces~\cite{naseem2022identification,zhang2021breaking,dym2020social}. Yet, less is known about how users themselves discuss privacy, interpret regulatory frameworks, and express emotional reactions to perceived risks. From a human-information interaction perspective~\cite{wilson1999models,fidel2012human,kuhlthau1991inside,kishnani2022privacy,noah2022privacy}, understanding how privacy concerns are articulated and negotiated within online communities is critical for modeling trust, self-disclosure, and perceived safety in information exchange~\cite{fell2020human,fisher2005theories,savolainen2023everyday,dev2018privacy,surani2022understanding}.

To address this gap, we analyze privacy-related discourse across $14$ subreddits spanning both \emph{mental-health} (e.g., \texttt{r/Depression}, \texttt{r/Anxiety}) and \emph{regulatory/privacy} domains (e.g., \texttt{r/GDPR}). This dual-domain design enables cross-context comparison between therapeutic and policy-oriented spaces, capturing how users negotiate emotional tone and information norms across divergent communicative settings~\cite{zimmer2020but,agosto2017don}. We collected $10{,}119$ posts and $65{,}385$ comments using a custom-built web scraper and applied lexicon-based sentiment analysis, supplemented with temporal and inferential statistical analyses, to examine how privacy discourse shapes community-wide emotion, temporal dynamics, and inter-community alignment within mental health information ecosystems. Our work offers three primary~\textbf{contributions}:

\begin{itemize}
    \item We identify key privacy-related themes emerging from Reddit mental health discourse, including concerns over data exposure, regulatory compliance, and institutional trust.
    
    \item We compare sentiment dynamics across mental health and regulatory subreddits, revealing strong emotional alignment within mental health communities (e.g., \texttt{r/Bipolar}–\texttt{r/ADHD}: $0.877$; \texttt{r/Anxiety}–\texttt{r/Depression}: $0.849$), contrasted with divergent affect in regulatory threads.
    
    \item We demonstrate that privacy-related discussions influence emotional tone and evolve over time, reflecting heightened awareness following major HIPAA and GDPR events. These findings contribute to a broader understanding of privacy discourse within online mental health information interaction and inform the design of privacy-aware digital support systems and data governance frameworks.
\end{itemize}

\section{Method}

\subsection{Data Collection and Cleaning}
We implemented a custom Reddit collector using the Python Reddit API Wrapper (PRAW) to retrieve public posts and comments from $2022$ to $2025$. For each item, we collected the identifier, title, score, comment count, timestamp, and body text. Data processing used \texttt{pandas} and \texttt{NumPy}~\cite{mckinney2012python}. We selected \(14\) subreddits spanning two comparative groups:
\begin{itemize}
    \item \textbf{Data Access-Related Subreddits:} We included  \texttt{r/GDPR} (950 posts), \texttt{r/hipaaviolations} (73 posts), \texttt{r/datasecurity} (105 posts), \texttt{r/HIPAA} (968 posts) . Posts were filtered for healthcare-related data breaches, regulatory compliance, and public awareness of privacy laws.

    \item \textbf{Mental Health Subreddits:} Mental health-focused subreddits were chosen for high engagement and reputation as supportive communities, including \texttt{r/mentalhealth} (1020 posts), \texttt{r/ADHD} (564 posts),  \texttt{r/Anxiety} (975 posts),    \texttt{r/bipolar} (846 posts),\texttt{r/depression} (885 posts),   \texttt{r/mentalillness} (958 posts),
    \texttt{r/therapy} (967 posts),
    \texttt{r/schizophrenia} (866 posts),
    \texttt{r/MentalHealthSupport} (934 posts), \texttt{r/Anxiety} (975 posts), \texttt{r/psychology} (8 posts). 
\end{itemize}

We analyzed only publicly available content, did not retain usernames or other direct identifiers, and report aggregate results. Data handling followed Reddit’s API terms. Given no interaction or intervention, the work was treated as non–human-subjects research. The final dataset comprised $10,119$ posts and $65,385$ comments.

\subsection{Analysis}

We conducted sentiment analysis that captured emotional tones across mental healthcare subreddits, extending our research beyond data privacy and security to understand user sentiment on mental health in online communities. To classify posts and comments as positive (score $>$ 0), negative (score $<$ 0), or neutral (score = 0), we used Python's TextBlob~\cite{loria2018textblob} with pre-trained models (chosen for interpretability). To analyze sentiment relationships between subreddits, we calculated cosine similarity (cosine similarity is standard in IR for distributional comparison~\cite{salton1983modern}) for sentiment distributions. To test whether sentiment scores differed significantly across discussion topics, we conducted a one-way ANOVA on posts categorized by four privacy/security themes: \emph{HIPAA}, \emph{GDPR}, \emph{Data Breach}, and \emph{Cyber-Threats}. Each category was derived using keyword dictionaries created from domain expertise and corpus inspection (e.g., ``medical records," ``HIPAA violation," ``phishing," ``ransomware," ``leak," ``compliance"), ensuring transparent, replicable topic tagging. Here, ``before" sentiment refers to posts and comments preceding the first appearance of a privacy-related keyword within a thread, while ``after" sentiment captures subsequent discussion following that exposure; exposure itself was operationalized using curated keyword dictionaries (e.g., HIPAA, GDPR, data breach terms). To mitigate sparsity and preserve anonymity, analyses were conducted at the post and thread level rather than modeling individual user trajectories. We further applied a Chi-Square test to examine whether privacy discussions were evenly distributed across subreddit categories and a Wilcoxon Signed Rank Test to compare sentiment before and after exposure to such discussions within threads, aggregating sentiment at the post level to preserve anonymity. Temporal trends were analyzed by grouping posts by month and year to track the evolution of privacy discussions from $2020$ to $2025$, revealing surges following major HIPAA or GDPR events. Given multiple hypothesis tests, we report significance cautiously and emphasize effect direction and magnitude alongside $p$-values.

\section{Results and Discussion}
\label{sec:results}

\subsection{Sentiment Analysis : Mental Health}
To understand the diversity of mental health experiences and concerns shared on Reddit, we analyzed $10,119$ posts and $65,385$ comments from both mental health-focused subreddits and data access and privacy-related subreddits. This analysis aimed to capture not only the range of user sentiment, spanning positive, negative, and neutral tones, but also how users engage with mental health and privacy topics on social media. These patterns echo Azzopardi et al.’s findings that online information sharing involves ongoing risk assessment~\cite{azzopardi2025assessing}, while highlighting how such risk talk is emotionally entangled in mental health communities.

\begin{table}[ht]
\centering
\small
\caption{Sentiment Distribution Within the Analyzed Reddit Dataset}
\label{tab:sentiment_distribution}
\begin{tabular}{p{2cm}rrrrr}
\toprule
\textbf{Keyword} & \textbf{Positive} & \textbf{Neutral} & \textbf{Negative} & \textbf{Total Posts} \\
\midrule
ADHD & 365 & 6 & 193 & 564 \\
Anxiety & 461 & 24 & 490 & 975 \\
GDPR & 640 & 92 & 218 & 950 \\
MentalHealthSupport & 501 & 21 & 412 & 934 \\
bipolar & 514 & 14 & 318 & 846 \\
datasecurity & 90 & 9 & 6 & 105 \\
depression & 430 & 36 & 419 & 885 \\
hipaa & 598 & 96 & 274 & 968 \\
hipaaviolations & 43 & 6 & 24 & 73 \\
mentalhealth & 510 & 71 & 439 & 1020 \\
mentalillness & 461 & 30 & 467 & 958 \\
psychology & 8 & 0 & 0 & 8 \\
schizophrenia & 449 & 100 & 317 & 866 \\
therapy & 623 & 19 & 325 & 967 \\
\bottomrule
\end{tabular}

\vspace{2pt}
{\footnotesize \textit{Note: Sentiment counts reflect the collected dataset and may not represent sentiment across Reddit as a whole.}}

\end{table}

\begin{figure}[htbp]
  \centering
  \includegraphics[
    width=\columnwidth,
    trim=0.25cm 0.35cm 0.25cm 0.20cm,
    clip
  ]{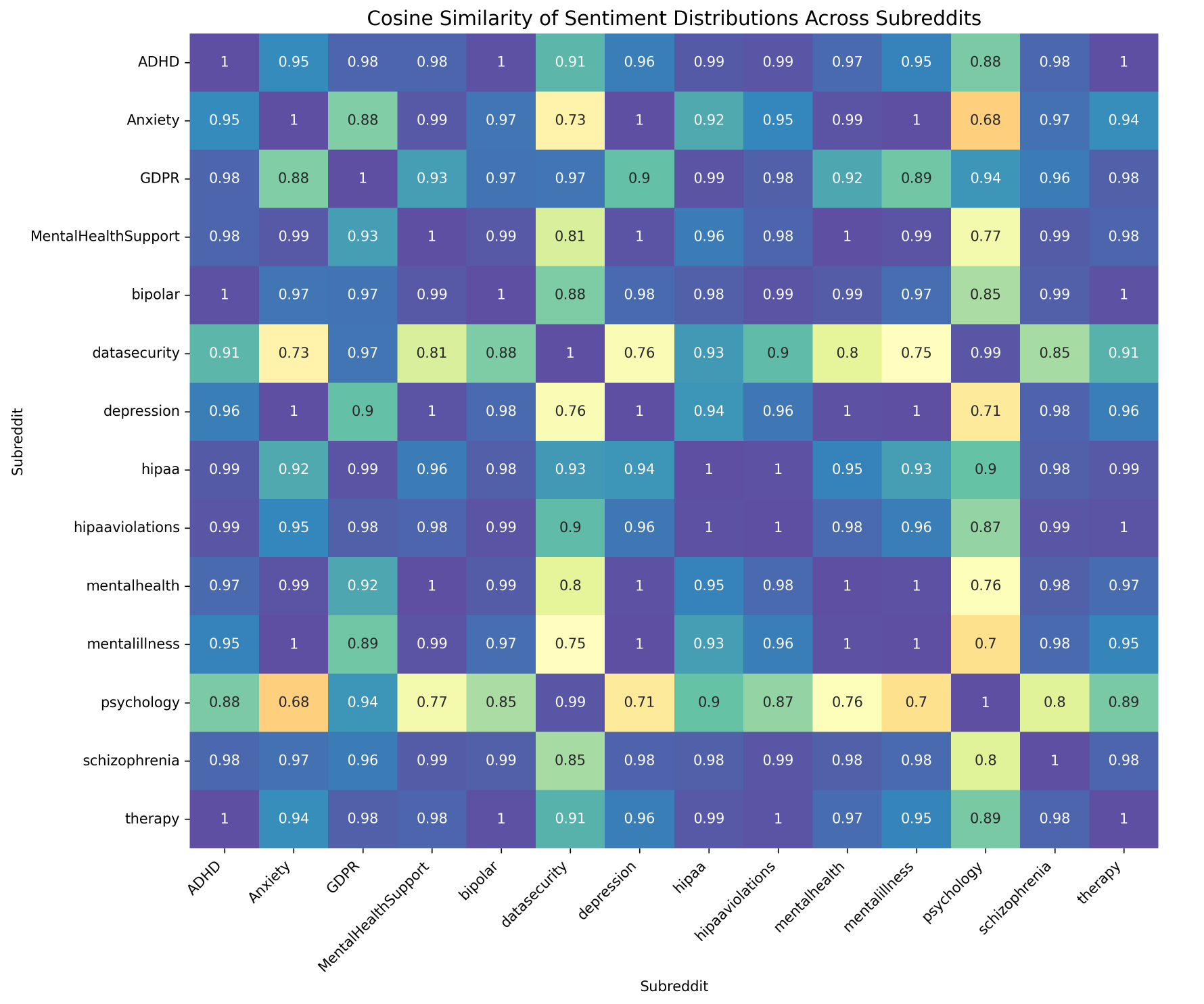}
  \caption{Heatmap Illustrating Similarity of Sentiment Patterns Across Mental Health Subreddits and Topics}
  \label{fig:heat}
  \Description{Heatmap showing cosine similarity of sentiment distributions across mental health and privacy subreddits, with most values indicating high similarity.}
\end{figure}

\subsection{Privacy-related Discussions}
We conducted an ANOVA one-way test to examine sentiment variations across different privacy-related discussions. The resulting p-value of \(1.7510 \times 10^{-6}\) is significantly below the threshold of \(0.05\), leading us to reject the null hypothesis. This confirms that sentiment scores differ significantly across these topics, emphasizing the varying emotional responses elicited by dtifferent privacy concerns. HIPAA-related discussions showed the lowest sentiment, with users citing healthcare privacy violations, unauthorized record access, and institutional negligence. Many posts highlighted concerns about inadequate encryption measures in electronic health records and failures in regulatory enforcement. The emotional charge in these discussions stemmed from personal experiences where users or their acquaintances suffered from medical data breaches, leading to distrust in healthcare institutions.

Conversely, discussions about data breaches recorded the highest sentiment scores, suggesting a more solution-oriented approach. Users often engaged in technical discourse on mitigation strategies, encryption standards, and best practices for securing sensitive information. The shift from fear to problem-solving within these conversations indicates that discussions surrounding  measures contribute to a sense of agency and control over data protection. GDPR and cyberattack-related discussions demonstrated a moderate sentiment distribution. GDPR topics remained largely neutral, as they primarily involved legal compliance, regulatory obligations, and corporate policies rather than deeply personal experiences. Cyberattack-related discussions exhibited a dual nature, where users expressed concern over emerging threats but also engaged in knowledge-sharing on defensive security practices, penetration testing, and threat modeling. Where the average sentiment score of data security tends to have a value of $0.08$ and where as the GDPR holds  a value slightly above $0.06$ and next comes the HIPAA where the sentiment score falls at $0.04$. 

We also conducted a Chi-Square test to evaluate whether privacy themes are evenly distributed across subreddit domains. The obtained p-value of \(0.03\) is below the significance threshold of \(0.05\), leading us to reject the null hypothesis. The Chi-Square statistic of \(5596.67\) with \(50\) degrees of freedom further confirms a structured variation in how privacy topics are discussed across different Reddit communities. Our analysis revealed that data breach-related discussions were predominantly concentrated in technology-focused subreddits. Legal and compliance subreddits, such as \texttt{r/GDPR} and \texttt{r/HIPAA}, exhibited a high prevalence of regulatory discussions, including corporate accountability and legal frameworks~\cite{andenaes1968legal}. Mental health-focused subreddits, such as \texttt{r/mentalhealthsupport} and \texttt{r/depression}, displayed sporadic engagement with privacy themes, mainly reacting to specific high-profile incidents. While this concentration is expected given HIPAA's healthcare-specific scope~\cite{das2022privacy,stevenson2025your}, we also observed HIPAA references in mental health support subreddits, where the term was often used to express broader concerns about confidentiality and trust rather than to discuss legal compliance explicitly.

\begin{figure}[htbp]\centering
\includegraphics[width=1\linewidth]{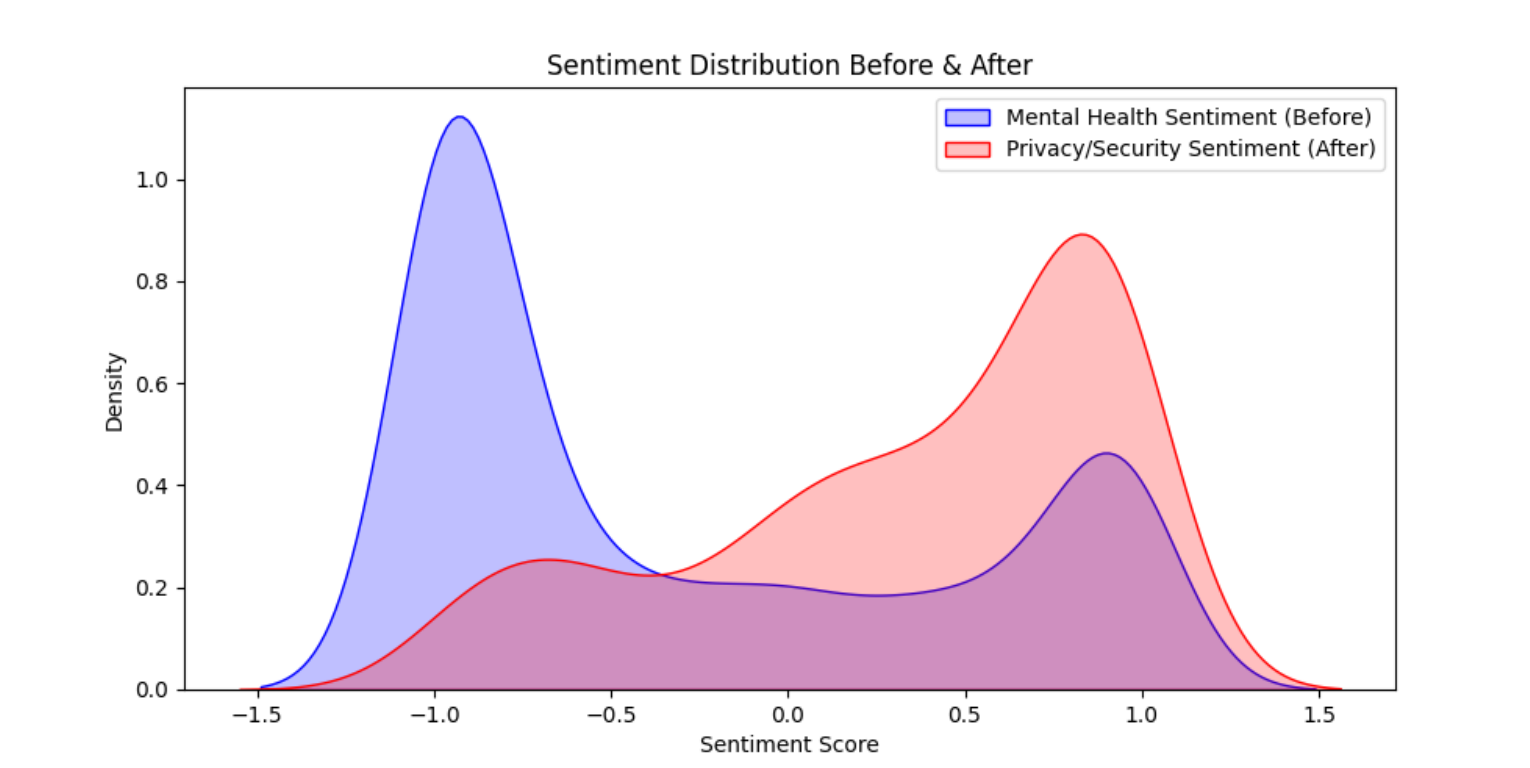}
\caption{Sentiment Distribution Before and After Privacy and Data Access Discussions (Wilcoxon Signed Rank Test)}
\Description{Density plot comparing sentiment before and after privacy discussions, showing a shift from negative to more positive sentiment.}
\label{fig:sentiment_shift}
\end{figure}

\subsection{Sentiment Shifts}
We analyzed sentiment shifts before and after engagement in privacy discussions using the Wilcoxon Signed Rank Test. The analysis revealed a statistically significant change in sentiment, indicating that participation in privacy conversations alters user perception. Before engaging in privacy discussions, sentiment scores were predominantly negative, especially in mental health-related subreddits. Users expressed distress over privacy violations, unauthorized data access, and the perceived lack of regulatory enforcement. Many posts detailed personal experiences of data misuse, contributing to heightened anxiety and distrust in digital platforms. Following engagement in privacy discussions, we observed a shift towards more neutral or slightly positive sentiment. Users reported feeling better informed about data protection strategies and best practices. Knowledge-sharing about encryption, secure browsing habits, and regulatory rights contributed to increased user confidence in managing their digital footprint. Figure~\ref{fig:sentiment_shift} illustrates these sentiment shifts, showing that privacy discussions contribute to reducing anxiety and promoting informed discourse around data protection.

\subsection{Emergence of Privacy Concerns}
Our time series analysis, conducted across all the reddits in general, indicates that between $2020$ and $2025$, discussions on privacy and mental health experienced a significant upward trend on Reddit. HIPAA-related posts increased steadily, appearing consistently in multiple months throughout $2020$ and $2021$. Similarly, data security discussions exhibited a rising trajectory, with an increasing number of posts per month from mid-$2020$ onwards. During the same period, mental health-related conversations, particularly in subreddits like r/depression, r/mentalhealth, and r/mentalillness, gained substantial traction. This rise in engagement coincided with the COVID-19 pandemic, suggesting that increased digital reliance heightened public awareness of both mental health and privacy concerns. Regulatory compliance-related discussions also intensified, with HIPAA violation reports experiencing a significant rise in $2021$, likely due to growing awareness of medical data security breaches. Similarly, data breach conversations continued their upward trend from $2021$ to $2022$, highlighting the increasing concern over digital privacy risks.

\begin{figure}[htbp]\centering
\includegraphics[width=1\linewidth]{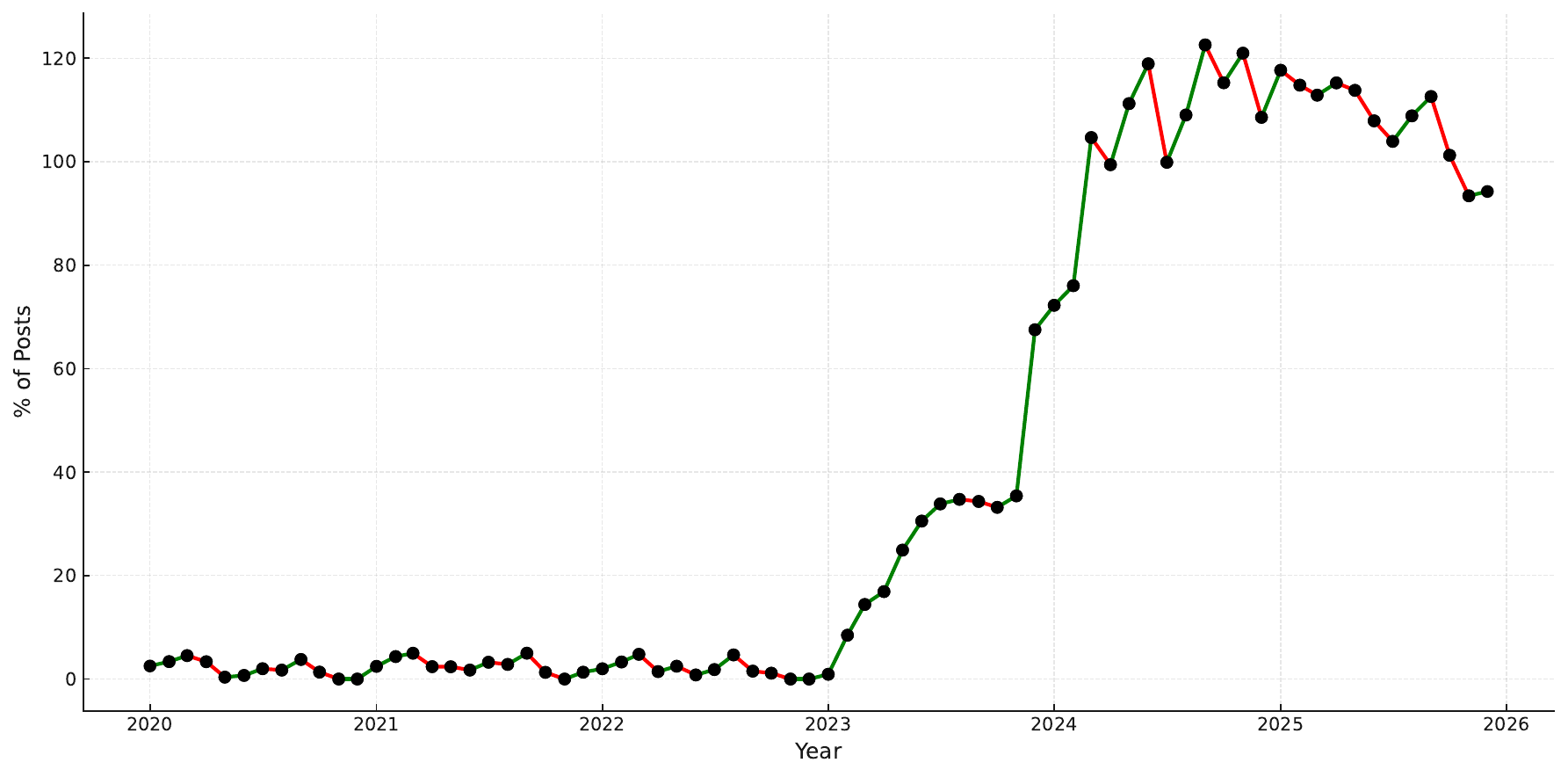}
\caption{Time Series Analysis Showing Monthly Discussion Trends in Privacy and Mental Health Discussions on Reddit}
\Description{Line chart showing monthly post trends on Reddit, with a sharp rise in privacy and mental health discussions after 2023.}
\label{fig:time_series}
\end{figure}

Starting in $2023$, privacy-related discussions exhibited an accelerated increase, driven by real-world events and evolving regulatory frameworks~\cite{skjong2009regulatory}. HIPAA-related posts peaked in April $2023$, comprising 10\% of total posts, with 7\% specifically addressing HIPAA violations. This suggests heightened public concern over medical data privacy breaches. GDPR discussions, in contrast, showed intermittent spikes, particularly in September and October $2024$, likely reflecting regulatory enforcement actions and policy updates. The recurring presence of HIPAA violation discussions throughout $2023$ indicates sustained public anxiety regarding the adequacy of privacy protections in healthcare institutions. Meanwhile, data access related discussions followed a more linear growth pattern and increased steadily without pronounced fluctuations. Unlike HIPAA and GDPR topics, which tended to peak in response to regulatory events, data breach concerns remained a persistent focus and reflected long-term public interest in threat mitigation strategies. Collectively, these trends point to a broader shift in digital discourse in which privacy has become an integral component of online mental health conversations. Figure~\ref{fig:time_series} illustrates monthly trends in both mental health and privacy-related discussions from $2020$ to $2025$, highlighting a pronounced post-$2023$ rise in privacy-related discourse alongside sustained growth in mental health discussions.

Our analysis show that, mental health communities such as Bipolar–ADHD, Anxiety–Depression, and MentalHealthSupport– MentalIllness exhibit highly similar sentiment distributions, while regulatory and policy-oriented subreddits maintain distinct affective profiles. Over the same period, privacy-focused discourse on HIPAA, GDPR, data exposure, and institutional trust increased by about 50\% from $2020$ to $2025$, with sharp spikes following major regulatory events. These patterns show that privacy concerns actively shape emotional dynamics and mental health information interaction by affecting how users frame experiences, what personal details they disclose, and whether and how they engage in support-seeking or advisory exchanges, consistent with prior work on privacy and security concerns in mental health and health-related settings~\cite{surani2023security,saka2025watch,joshi2020substituting,das2020change}. When privacy-related information becomes salient, users recalibrate trust and perceived safety through sensemaking, which then changes downstream behaviors such as seeking advice, offering support, limiting interaction depth, or withholding sensitive details. These findings argue for information systems that surface privacy cues~\cite{adhikaripolicypulse,das2020humans,das2019privacy} in ways that support informed sharing decisions and invite further research on how such cues influence disclosure thresholds, interaction depth, and long-term participation in sensitive online environments.

\section{Limitations and Future Work}
\label{futureworklimitations}
We could not access posts that were deleted by users or removed by moderators, which may bias topic and sentiment distributions, and private or restricted subreddits were out of scope, limiting the diversity of observed discourse. Future work will incorporate crowdsourced annotation frameworks~\cite{ide2004international} to expand sentiment and privacy-related labels and report inter-annotator agreement (e.g., Fleiss’ $\kappa$, Cohen’s $\kappa$). We also plan platform-level analyses of encryption, access control, API behavior, and regulatory compliance using tools such as OWASP ZAP and Burp Suite to identify vulnerabilities relevant to digital mental-health contexts.
\section{Conclusion}
\label{sec:conclusion}
We conducted a large-scale analysis of privacy concerns in online mental health discussions on Reddit, examining $10{,}119$ posts and $65{,}385$ comments from $14$ subreddits. Our sentiment analysis showed strong emotional alignment across mental health communities, including a cosine similarity of $0.994$ between \texttt{r/MentalHealth} and \texttt{r/MentalHealthSupport}. Time series analysis indicated a steady rise in privacy-related discussions from $2013$ to $2025$, with HIPAA-related conversations peaking at 12\% of total posts in late $2023$ and GDPR discussions spiking in Q3 $2024$, likely reflecting regulatory updates. A Chi-Square test confirmed that privacy and data breach topics are not uniformly distributed across subreddit domains ($\chi^2=5596.67$, $p=0.03$, $df=50$), with data breach discussions concentrated in technical forums and HIPAA and GDPR concerns more prevalent in healthcare-related subreddits. A one-way ANOVA further showed significant sentiment differences across privacy themes ($p=1.7510\times10^{-6}$), with the greatest sentiment variation in data breach topics and the lowest sentiment in HIPAA compliance discussions. Our findings highlight the critical intersection of mental health discourse and data privacy in online forums.

\section{Acknowledgment}
We would like to acknowledge the Das Agency and Security (DAS) Lab at George Mason University and the Inclusive Security and Privacy-focused Innovative Research in Information Technology (INSPIRIT) Lab at the University of Denver. The opinions expressed in this work are solely those of the authors.

\bibliographystyle{ACM-Reference-Format}
\balance
\bibliography{CHIIR2026}
\end{document}